\documentclass[pre,aps,twocolumn,showpacs,showkeys,reprint,superscriptaddress,amsmath,amssymb]{revtex4-1}

\usepackage{graphicx}
\usepackage{amssymb}
\usepackage{color,soul}
\usepackage{amsthm}

\usepackage[hyperindex=true,pdftitle={The Quantum Kuramoto Model},
colorlinks=true,pagebackref=false,citecolor=blue,plainpages=false,pdfpagelabels,
linkcolor=blue,urlcolor=blue]{hyperref}

\begin{document}

\title{Synchronization in a semiclassical Kuramoto model}

\author{Ignacio Hermoso de Mendoza}
\affiliation{Departamento de F\'{\i}sica de la Materia Condensada, Universidad
de Zaragoza, E-50009 Zaragoza, Spain.}

\author{Leonardo A. Pach\'on}
\affiliation{Grupo de F\'isica At\'omica y Molecular, Instituto de F\'{\i}sica,  Facultad de Ciencias Exactas y Naturales, 
Universidad de Antioquia UdeA; Calle 70 No. 52-21, Medell\'in, Colombia}

\author{Jes\'us G\'omez-Garde\~nes}
\affiliation{Instituto de Biocomputaci\'on y F\'{\i}sica de Sistemas Complejos,
Universidad de Zaragoza, E-50018 Zaragoza, Spain.}
\affiliation{Departamento de F\'{\i}sica de la Materia Condensada, Universidad
de Zaragoza, E-50009 Zaragoza, Spain.}

\author{David Zueco}
\affiliation{Departamento de F\'{\i}sica de la Materia Condensada, Universidad
de Zaragoza, E-50009 Zaragoza, Spain.}
\affiliation{Instituto de Ciencia de Materiales de Arag\'on (ICMA),
  CSIC-Universidad de Zaragoza, E-50012 Zaragoza, Spain.}
\affiliation{Fundaci\' on ARAID, Paseo Mar\'{\i}a Agust\'{\i}n 36, E-50004
Zaragoza, Spain. }

\date{\today}
\begin{abstract}
Synchronization is an ubiquitous phenomenon occurring in social, biological and technological systems when the internal rhythms of a large number of units evolve coupled. 
This natural tendency towards dynamical consensus has spurred a large body of theoretical and experimental research during the last decades. 
The Kuramoto model constitutes the most studied and paradigmatic framework to study  synchronization. 
In particular, it shows how synchronization shows up as a phase transition from a dynamically disordered state at some critical value for the coupling strength between the interacting units. 
The critical properties of the synchronization transition of this model have been widely studied and many variants of its formulations has been considered to address different physical realizations. 
However, the Kuramoto model has been only studied within the domain of classical dynamics, thus neglecting its applications for the study of quantum synchronization phenomena. 
Based on a system-bath approach and within the Feynman path-integral formalism, we derive the equations for the Kuramoto model by taking into account the first
quantum fluctuations.
We also analyze its critical properties being the main result the derivation of the value for the synchronization onset. 
This critical coupling turns up to increase its value as quantumness increases, as a consequence of the possibility of tunnelling that quantum fluctuations provide.
\end{abstract}

\maketitle

\section{Introduction}

Synchronization is perhaps the  most cross-disciplinary concept of emergence of collective behavior \cite{Strogatz2004} as it is manifested across many branches of natural and social sciences. 
Ensembles of neurons, fireflies or humans are prone to synchronize their internal rhythms when they become coupled enough, producing a macroscopic dynamically coherent state. In all these seemingly unrelated situations, no matter the precise nature of the coupled units, interaction drives system's components to behave homogeneously. 
Thus, the study about the microscopic rules that drive ensembles towards synchrony has a long and fruitful history since the seminal observations made by Christiaan Huygens \cite{Pikovsky2003,Manrubia2004,Boccaletti2008}.

The mathematical formulation of the first models showing
synchronization phenomena dates back to the $70$'s when, after some
preliminary works by Peskin and Winfree \cite{Strogatz2000}, Kuramoto \cite{Kuramoto75} formalized his celebrated model. The Kuramoto model incorporates the minimum dynamical ingredients aimed at capturing a variety of physical phenomena related with the onset of synchronization. In particular, the Kuramoto model links physical concepts such as self-organization, emergence, order in time and phase transitions, thus revealing as the most paradigmatic framework to study synchronization \cite{Strogatz2000, Nadis2003,Bonilla2005}.

\begin{figure*}
\centering
\includegraphics[width=0.90\textwidth]{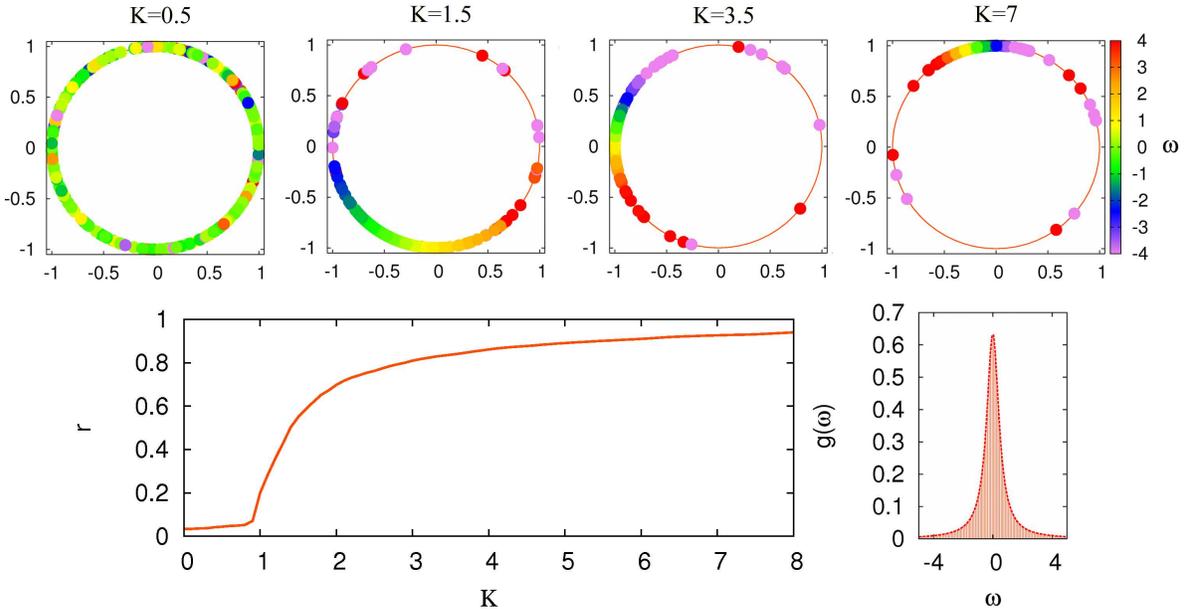}
\caption{(color online) {\bf Synchronization in the classical Kuramoto model}. 
Each panel on the top shows the collection of oscillators situated in the unit circle 
(when each oscillator $j$ is represented as ${\mbox e}^{{\mbox i}\theta_j(t)}$). 
The color of each oscillator represents its natural frequency. 
From left to right we observe how oscillators start to concentrate as the coupling $K$ increases. 
In the panels below we show the synchronization diagram, {\it i.e.}, the Kuramoto order parameter $r$ as a function of $K$. 
It is clear that $K_{\mathrm c}=1$ as obtained by using the distribution $g(\omega)$ shown in the right panel.}
\label{fig:1}
\end{figure*}

Despite the large body of literature devoted to the Kuramoto model and
its variants, its study has always been restricted to the classical
domain.  At first sight, given the usual nature (scale) of the systems
in which synchronization is typically observed, it seems superflous
thinking of a quantum theory for the Kuramoto model. However, there is
not doubt about the fundamental importance of studying quantum
fluctuations within the emergence of synchronized states
\cite{Goychuk2006, Zhirov2008, Mari2013, Liu2013, Giorgi2012, Manzano2013, Lee2013,
Lee2014}. 
Moreover, the Kuramoto model has been implemented on circuits and micro and nanomechanical structures \cite{TM&98,Matheny2013}, 
systems which have already met the quantum domain \cite{LaHaye2004, OConnell2010}.
At the quantum level, synchronization, understood as the emergence of a coherent behaviour from an incoherent situation 
in the absence of external fields, is reminiscent of the phenomena such as condensation of Bose-Einsten and
has been observed in interacting condensates of quasiparticles \cite{BL&08,CD&13}.
Additionally, synchronization has been suggested to occur in ensamble of atoms and 
enhance the coherence time next generation of lasers \cite{XT&13}
Thus, moved by its fundamental and applied importance, in this work we provide the semiclassical version of the Kuramoto model in an attempt for understanding the influence that quantumness has on the emergence of synchronized states. 

Our work in this paper consists, as stated by Caldeira and Leggett \cite{Caldeira1983}, on finding consistent equations that in the classical limit matches the Kuramoto model. 
Our derivation relies on the quantization of open systems in the framework of Feynman's  path-integral formalism. 
We compute the first quantum corrections to the {\it classical}
Kuramoto model.   We also analyze its critical properties by deriving the critical point from which synchronization shows up and determine how quantum fluctuations affect this synchronization transition. 

The rest of the paper is organized as follows.  In the next section
\ref{sec:classical} we review the main features of the classical
model.   
Section \ref{sec:q-K}, constitutes the main part of our work, there
we present the semiclassical equations and draw our numerical results on the
sync dynamics.  In section \ref{sec:K} we derive the critical value
for the synchronization transition.  We write our conclusions on
\ref{sec:conc}, sending most of the technical steps for the
semiclassical calculations and the critical value to the Appendices.


\section{The classical Kuramoto model.}
\label{sec:classical}

The original Kuramoto model \cite{Kuramoto75} considers a collection of $N$ phase-oscillators, {\it i.e.}, it assumes that the characteristic time scale of their amplitudes is much faster than that for the phases. Thus, the dynamical state of the $i$-th unit is described by an angular variable $\theta_i\in(0,2\pi]$ whose time evolution is given by: 
\begin{equation}
\label{K0}
\dot \theta_i = \omega_i + \frac{K}{N} \sum_{j=1}^N \sin (\theta_i - \theta_j)\; .
\end{equation}
The above equation thus describes a set of weakly coupled phase-oscillators whose internal 
(natural) frequencies $\{\omega_i\}$ are, in principle, different as they are assigned following a frequency distribution 
$g(\omega)$ that is assumed to be uni-modal and even around the mean frequency $\Omega$ of the population, 
$g(\Omega+\omega)=g(\Omega-\omega)$.  

In the uncoupled limit ($K=0$) each element $i$ describes limit-cycle oscillations with characteristic frequency $\omega_i$. 
Kuramoto showed that, by increasing the coupling $K$ the system experiences a transition towards complete synchronization, {\it i.e.},  a dynamical state in which $\theta_i(t)=
\theta_j(t)$ $\forall i, j$ and $\forall t$. This transition shows up when the coupling strength exceeds a critical value whose exact value is:
\begin{equation}
\label{crit-clas}
 K_{\mathrm c}=\frac{2}{\pi g(\Omega)}\;.
\end{equation} 

To monitor the transition towards synchronization, Kuramoto introduce a complex order parameter:
\begin{equation}
\label{Kuraparam}
r(t){\mbox e}^{{\mbox i}\Psi(t)}=\frac{1}{N}\sum_{j=1}^N {\mbox
  e}^{{\mbox i}\theta_j(t)} \;.
\end{equation}
The modulus of the above order parameter, $r(t)\in[0,1]$, measures the coherence of the collective motion, reaching the value $r=1$ when the system is fully synchronized, while $r=0$ for the incoherent solution. On the other hand, the value of $\Psi(t)$ accounts for the average phase of the collective dynamics of the system.

In Figure \ref{fig:1} we have illustrated the synchronization in the Kuramoto model. 
The panels in the top show, for different values of 
the coupling $K$, how the oscillators concentrate as $K$ increases. 
Below we have shown the usual synchronization diagram  $ r(K)$ for which the exact 
value of $r$  for each $K$ is the result of a time average of $r(t)$ over a large enough 
time window. 
In this diagram we can observe that $K_{\mathrm c}=1$ as a result of using the distribution 
$g(\omega)$ shown in the right.

Let us note that the all-to-all coupling considered originally by Kuramoto can be trivially 
generalized to any connectivity structure by introducing the coupling matrix $A_{ij}$ inside 
the sum in Eq.~(\ref{K0}) so that each term $j$ accounting for the interaction between oscillator 
$i$ and $j$ is assigned a different weight. 
The latter allows for the study of the synchronization properties of a variety of real-world 
systems for which interactions between constituents are better described as a complex 
network  \cite{Boccaletti2006}.
The formalism developed in this work is fully general and valid for
any  form of $K_{ij}$ thus making possible the extension of the large
number of studies about the Kuramoto model in any topology \cite{PRsync} to the semiclassical domain. 
However, the numerical part of our work will deal with the all-to-all coupling for the sake of comparison with the original Kuramoto work.


\section{Quantization of the Kuramoto model.} 
\label{sec:q-K}

The most important problem when facing the quantization of the Kuramoto model is its 
non-Hamiltonian character since, as introduced above, equation (\ref{K0}) assumes 
the steady-state for the dynamical state of the amplitude of the oscillators. 
Thus, a question arises, how do we
introduce quantum fluctuations in the Kuramoto model?
One possible choice is to resort to the original microscopic dynamics of amplitude and 
phases and then identify the underlying Hamiltonian dynamics. 
However, many different dynamical setups can have the Kuramoto model as their corresponding 
limiting case of fast amplitude dynamics. 
Thus, in order to keep the flavor of generality of the Kuramoto model, 
it is desirable not to resort to 
any specific situation (Hamiltonian) and introduce quantum fluctuations directly.

A similar problem was faced by Caldeira and Leggett in the eighties \cite{Caldeira1983} 
when they studied the influence of dissipation in quantum tunneling. 
In their case, the corresponding classical dynamics dates back to the studies on activation theory 
by Kramers \cite{Hanggi1990}. 
Classically, a particle in a potential experience an energy barrier to surmount, that is typically 
acquired from thermal fluctuations. 
On the other hand, a quantum  particle finds in tunnelling an alternative way to bypass an energy 
barrier. 
Caldeira and Leggett were thus interested in quantifying the catalytic effect of tunnelling in (effectively) 
lowering the energy barriers. 
However, as in the Kuramoto model, Kramers activation theory is based in Langevin equations, {\it i.e.} stochastic
equations that are not directly obtained from any Lagrangian. Furthermore, most of reaction rate equations were phenomenological. 
Therefore, they searched for a consistent way for introducing quantum fluctuations regardless of the microscopic origin of the {\it  effective} classical evolution. As a byproduct their work opened the field of quantum Brownian motion in the most general way.

We take here the same route followed by Caldeira and Leggett to introduce quantum fluctuations in the Kuramoto model. In order to  
accomodate our dynamical system (\ref{K0}) to the framework provided in  \cite{Caldeira1983} we start by writting its corresponding Langevin equation:
\begin{equation}
\label{langevin}
\dot \theta_i = - \frac{\partial V}{\partial \theta_i} + \xi_i\;,
\end{equation}
with 
\begin{equation}
\label{eff-pot}
V (\theta_1, ..., \theta_N)  \equiv  - \sum_i \omega_i \theta_i + \frac{K}{N}\sum _{i,j} \cos (\theta_i - \theta_j)\;.
\end{equation}
As usual, $\xi_i$ is a Markovian stochastic fluctuating force with $\langle \xi_i (t) \rangle = 0$
and $\langle \xi_i (t) \xi_j (t^\prime) \rangle = 2 \delta_{ij} D\delta (t - t^{\prime})$. 
In the limit $D \to 0$, equation (\ref{langevin}) reduces to the Kuramoto model in equation (\ref{K0}).

Equation (\ref{langevin}) is nothing but a Langevin equation in the overdamped limit. 
It is first rather than second order in time as the inertia term is neglected. 
Consequently, the Kuramoto model can be viewed as a set of phases evolving in the overdamped limit.  
The absence of fluctuations in the limit $D\to 0$ means that the system of phases is 
at zero temperature, $D \sim T$. 
Such identification with a Langevin equation has been already used for generalizations 
of the original Kuramoto model taking into account noise and/or inertial effects \cite{Bonilla2005}. 
In particular, in \cite{Sakaguchi1988} it is shown that the critical value $K_{\mathrm c}$ reads:
\begin{equation}
\label{KCS}
K_{\mathrm c} = \frac{2}{\int_{-\infty}^{\infty}  {\rm d} \omega \frac{D }{D^2+\omega^2 } \, g(\omega) }\;,
\end{equation}
which, in the limit $D\to 0$, recovers the Kuramoto critical coupling (\ref{crit-clas}).

The key point of deriving the Langevin equation (\ref{langevin}) corresponding to the 
Kuramoto model is that it can be obtained from a fully Hamiltonian framework by coupling 
the system, in our case the coupled phases $\theta_i$, to a macroscopic bath or reservoir \cite{Hanggi1990}. 
In this way, both the damping and fluctuations are seen to be caused by the coupling of 
the system of phases to the bath. 
The Hamiltonian description is properly casted in the system-bath approach:
\begin{equation}
\label{Ht}
H_{\rm tot} = H_{\rm sys} + H_{\rm bath} + H_{\rm int}\;,
\end{equation}
where the bath is an infinite collection of harmonic oscillators with frequencies $\{\omega_{\alpha}\}$ 
(note that greek subindexes will denote the oscillators in the bath).  
In the case we are dealing with the {\it total} Hamiltonian reads:
\begin{equation}
\label{K-Ht}
H_{\rm tot}=\sum_i \frac{ \pi_i^2}{2}+V (\theta_1, ..., \theta_N)
+\frac{1}{2}\sum_{i, \alpha} P_{i, \alpha}^2 + \omega_\alpha^2 (Q_{i,\alpha} - \lambda_\alpha \theta_i)^2\;,
\end{equation}
where $\{(\theta_i, \pi_i)\}$ and $\{(Q_{i, \alpha} ,  P_{i,  \alpha})\}$ denote the  system and 
bath canonical coordinates, respectively, while $\lambda_{\alpha}$ stands for the coupling 
constant between bath and system coordinates. 

Under well defined conditions, the equations of motion for the system coordinates derived 
from the Hamiltonian (\ref{K-Ht}) lead to the the afore-derived overdamped Langevin equation 
(\ref{langevin}). 
In particular, one needs to assume: {\it (i)} thermalized initial conditions for the bath:
\begin{eqnarray} 
\langle Q_{i, \alpha} Q_{i^\prime, \alpha^\prime} \rangle &=&\delta_{i, i^\prime} \delta_{\alpha, \alpha^{\prime}} \; k_{\rm B}
T / \omega_ \alpha^2\;,
\\
\langle P_{i, \alpha} P_{i^\prime, \alpha^\prime} \rangle &=&\delta_{i, i^\prime} \delta_{\alpha, \alpha^{\prime}} \; k_{\rm B}
T\;,
\end{eqnarray} 
{\it (ii)} the frequency spectrum of the bath oscillators is flat (this assumption 
leads to the widely used Ohmic dissipation), and finally {\it (iii)} the changes in time of the velocity 
(acceleration) induced by the energy potentials are far slower than the energy loss induced by the 
coupling between the system and the bath (this is the situation when the system and the bath are 
strongly coupled) so that we could neglect the inertial term.


\subsection{The semiclassical equation}

Once we have a Hamiltonian description for the Kuramoto equation (\ref{K0}), we are ready to perform 
its quantization.  
First, we associate the phases and their associated momenta together with the positions and momenta 
for the bath by providing them with the canonical commutation rules.  
The hardest work is to find an effective quantum evolution depending only on phase operators, {\it i.e.} 
the so-called quantum Langevin equation. 
It turns out that such operators equation is a non-local in time differential equation, which makes it extremely 
difficult to manipulate in general. 
However, the quantum version of equation (\ref{langevin}) in the overdamped limit  is a $c$-number 
local differential equation \cite{Ankerhold2001,Ankerhold2005,
  Maier2010, Machura2004,uczka2005, Machura2006}. 
The full derivation for the quantum Langevin equation is based on the
Path Integral formulation.  It is lengthy and rather technical. 
Let us first present the final result (below), then a sketch of the
deriviation. Further details can be found in Appendix \ref{app:q-K}. 

The resulting quantum evolution in the Ito representation reads as follows:
 \begin{equation}
\label{q-l}
\dot \theta_i
= -\frac{V'_i}{F_i}
+\frac{\Lambda}{F_i}\sum_j\left(\beta V^\prime_j V^{\prime
    \prime}_{ij}-V^{\prime \prime \prime}_{jji}\right)
-\frac{\Lambda}{2F_i} V^{\prime \prime \prime}_{iii} + \sqrt {
  \frac{1}{F_i}}
\cdot\xi_i\;,
\end{equation}
where we have used the compact notation $V_{i, ..., k}^{\prime
  ... \prime} \equiv \partial_{\theta_i, ..., \theta_k} V$, 
$\xi_i$ is an stochastic force with the same statistics as in
(\ref{langevin}), 
\begin{equation}
F_i = {\rm e}^{-\frac{\Lambda}{D}  V^{\prime \prime}_{i i} }\;
\end{equation}
and $\Lambda$ is the \emph{quantumness} parameter:
\begin{align}
\Lambda = &\frac{2}{ m \beta} \sum_n \frac{1}{\nu_n^2 + \gamma \nu_n}
\\ \nonumber
=&
\frac{\hbar}{m \pi \gamma}
\left ( \Psi \left [
\frac{\hbar \beta \gamma}{2 \pi}
\right]
-\mathrm{C} + \frac{2 \pi}{\hbar \beta \gamma} \right ),
\end{align}
being $\mathrm{C}= 0.577...$ the Euler-Mascheroni constant and $\Psi$
the Digamma function.  
Note that in the limit $\hbar \beta  \gamma \to 0$  $\Lambda \to 0$.
Making $\Lambda \to 0$  the quantum Langevin \eqref{q-l} reduces to
the classical \eqref{langevin}.
This is a remarkable property.  Our result is perturbative in
$ \beta \Lambda$, giving the first quantum corrections containing, as a
limit, the Kuramoto model.
We notice that, being perturbative, $\beta \Lambda$ must be \emph {small},
wich means that our equation is valid at high temperatures and
damping.
As a drawback of the perturbative character, the model can not be
pushed to the zero temperature limit.  
Compared to its classical counterpart [$\beta \Lambda \to 0$, Eq. (\ref{langevin})],
Eq.~(\ref{q-l})  has a renormalized effective potential  (\ref{eff-pot})
 (third term in the r.h.s).   Besides,  both the
diffusion and consequently its noise terms are also modified by the
quantum fluctuations
(second and last terms in the r.h.s respectively). 

The noise, because of the $\sqrt {1 /
  F_i}$ is now multiplicative. 
In the limit $ \beta \Lambda \rightarrow 0$, 
 $F \to 1 $.
 Hence, in the classical limit the multiplicative noise switches
into additive noise. This immediately suggests that the multiplicative
nature is related to the underlying quantum stochastic
process. 
Quantum noise depends, undoubtedly, on the state of the
system, the dynamics of observables depend on the state the system
is and therefore, quantum noise in a Langevin equation must depend
upon the dynamics of the system itself. This explains the
multiplicative character of the noise in Eq.~\eqref{q-l} at the single
variable level. This result is consistent with previous works along
this line (see, e.g., Ref.~\cite{Machura2004,uczka2005,
 Ankerhold2005, Machura2006}).


\subsubsection{Sketch for  the derivation of Eq.~\eqref{q-l}}
 \label{sect:sketch} 
Any Langevin equation, classical or quantum, is an effective evolution
for the system of interest degrees of freedom. 
If we start with the total Hamiltonian, the bath degrees of freedom
need to be integrated out.  
In the quantum regime, this means taking the partial trace over
the bath Hilbert space.  
We follow here the program explained in Refs
\onlinecite{Ankerhold2001,Ankerhold2005}. 
The steps are as follows. \emph{i)} The equilibrium reduced density
matrix \cite{Weiss}:
\begin{equation}
\label{P}
\varrho_\beta ( {\bf \theta ; \theta^\prime})
=
\int {\rm d Q}^n 
\;
W_\beta ( {\bf Q, \theta ; Q^\prime, \theta^\prime })
\end{equation}
 is obtained in the overdamped limit.   Both ${\bf Q}$ and ${\bf
   \theta}$ are a shorthand notation for denote the bath ($Q_1,
 ...$) and system ($\theta_1, ..., \theta_N$) coordinates. 
In such a regime the damping is sufficiently strong to suppress 
the non-diagonal elements, \textit{coherences}, of the reduced density matrix, 
i.e., a regime where 
$\langle \theta_1, ..., \theta_N | \varrho_\beta | \theta_1^\prime,
..., \theta_N^\prime \rangle \sim \prod \delta (\theta_i - \theta_i^\prime)$.
We define
\begin{equation}
\label{P-def}
P_\beta ({\bf \theta}) := \varrho_\beta ( {\bf \theta ;
  \theta})
\, .
\end{equation}
As detailed in Appendix \ref{app:q-K},  the reduced density matrix in
the overdamped limit can be written as:
\begin{equation}
\label{rho-eq}
 P_\beta (\theta)
=
\mathcal{Z}^{-1}
{\rm e} ^{ -\beta \Lambda  \sum_i 
V_{i,i}^{\prime \prime}
}
\,
{\rm e}^{-\beta V +
\frac{1}{2}
 \beta^2 \Lambda \sum_i (
V_i^\prime
)^2 }
\, .
\end{equation}

Once the equilibrium density matrix is obtained, \emph{ii)} the
master equation for the probability distribution $P (q , t)$ [Cf. Eq.
\eqref{P-def}] is proposed.  Taking into account the results for the  harmonic
oscillator \cite{Maier2010} and the single particle case
\cite{Ankerhold2001, Machura2004}, the master equation can be formally
written as:
\begin{equation}
\label{qme-formal}
\partial_t P (\theta ; t) = \partial_\theta L \, P (\theta; t).
\end{equation}
The \emph{iii)} actual  master equation takes a Fokker
Planck form.  It  is by
obtained imposing that the equilibrium density distribution $P_\beta$
given by Eq.~\eqref{rho-eq} is stationary under \eqref{qme-formal}, $L \,
P_\beta (q ) = 0$.  
The final result is
\begin{align}
\label{qme-final}
\nonumber
\partial_tP 
= \sum_i\frac{{\partial}}{{\partial}{\theta}_i}
\Big\lbrace  
\Big[ 
&
\frac{V'_i}{{\Gamma}F_i}-\frac{\beta}{{\Gamma}F_i}{\Lambda}
{\sum_j}
V_j^\prime \, V_{i, j}^{\prime \prime}
 +\frac{\Lambda}{{\Gamma}F_i}
\sum_{j\neq{i}}V^{\prime\prime\prime}_{jji}\Big]  
\\ & + \frac{{\partial}}{{\partial}{\theta}_i}\left[ 
\frac{D}{\gamma^2F_i}\right] \Big\rbrace P.
\end{align}
Finally, \emph{iv)}  the Langevin equation \eqref{q-l} is obtained via
the equivalence of Fokker-Planck equations, as Eq.~\eqref{qme-final} and
Langevin-type equations
\cite{Garcia-Palacios2007}.

\section{The transition to synchronization in the semiclassical model.} 
\label{sec:K}

\begin{figure}
\centering
\includegraphics[width=0.95\columnwidth]{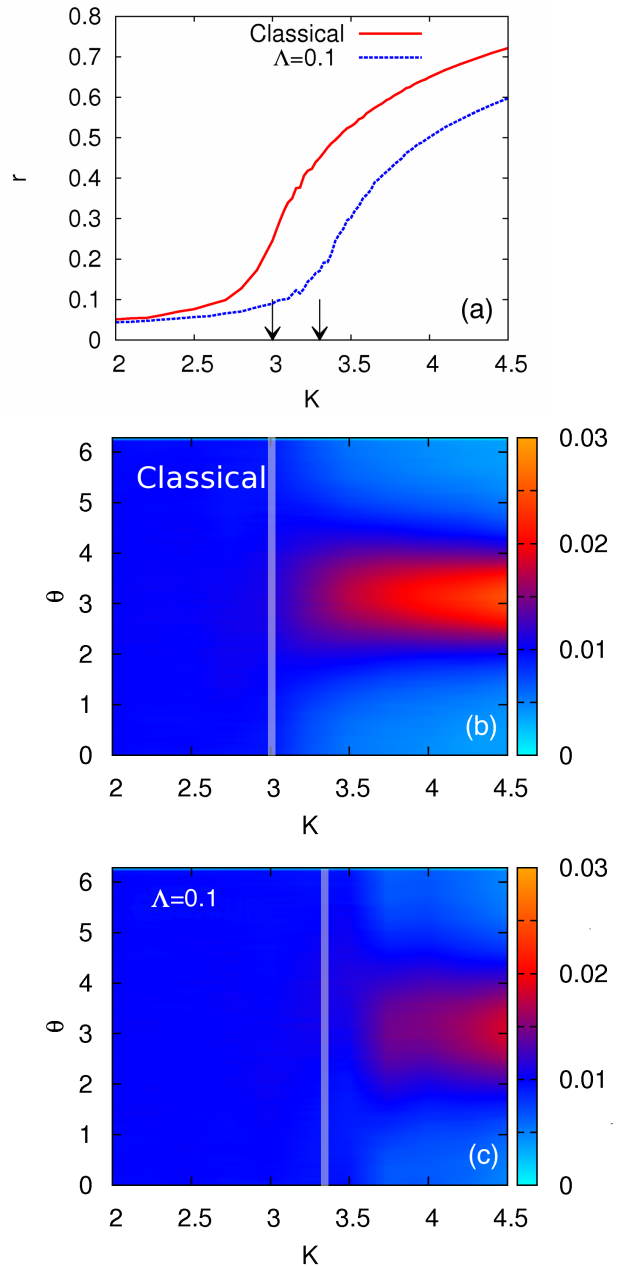}
\caption{(color online) {\bf Classical {\it vs.} Semiclassical synchronization transitions}. 
Panel  (a) shows the synchronization diagrams $r(K)$ for the classical 
($\Lambda=0$) and the quantum ($\Lambda=0.1$) Kuramoto models. 
In both cases the thermal noise is chosen such that $D=1$. 
The number of oscillators is $N=10^3$ and the distribution of natural  frequencies is given
in Eq.~(\ref{equ:Lorentzian}) centered in $\omega_0 = 0$ and $\alpha = 0.5$.
It is clear that the synchronization onset is delayed as soon as
quantumness enters  into play. 
In panels (b) and (c) we show the probability $P(\theta)$ of finding
an oscillator
at a given phase $\theta$ as a function of $K$. 
Note that for each value of $K$, the phases has been equally shifted
so that the mean phase is located at $\theta=\pi$. 
A thick grey line indicates the critical values $K_{\mathrm c}$ and $K_{\mathrm c}^{\mathrm q}$ 
for classical and quantum dynamics, respectively.
}
\label{fig:2}
\end{figure}
Once we  derived the semiclassical version of the Kuramoto equation,
it is natural to unveil the effects that quantum fluctuations induce in the 
transition to synchronization. 
As introduced previously, to study the synchronization transition one resorts 
to the order parameter $r$ [introduced in equation (\ref{q-l})] that reveals 
the synchronized state of the system. 
We solve both the classical Kuramoto model ($\Lambda = 0$) and the quantum 
one ($\Lambda > 0$) numerically, extracting from the dynamics the stationary value of $r$.
Through this work, the numerical calculations are performed with $N=10^3$ 
oscillators and the distribution of natural frequencies is Lorentzian: 
\begin{equation}
\label{equ:Lorentzian}
g(\omega  ; \omega_0, \alpha) = \frac{1}{\pi} \frac{\alpha}{(\omega - \omega_0)^2 + \alpha^2}\;,
\end{equation} 
with $\alpha = 0.5$ and centered around $\omega_0 = 0$.

\begin{figure*}
\centering
\includegraphics[width=0.85\textwidth]{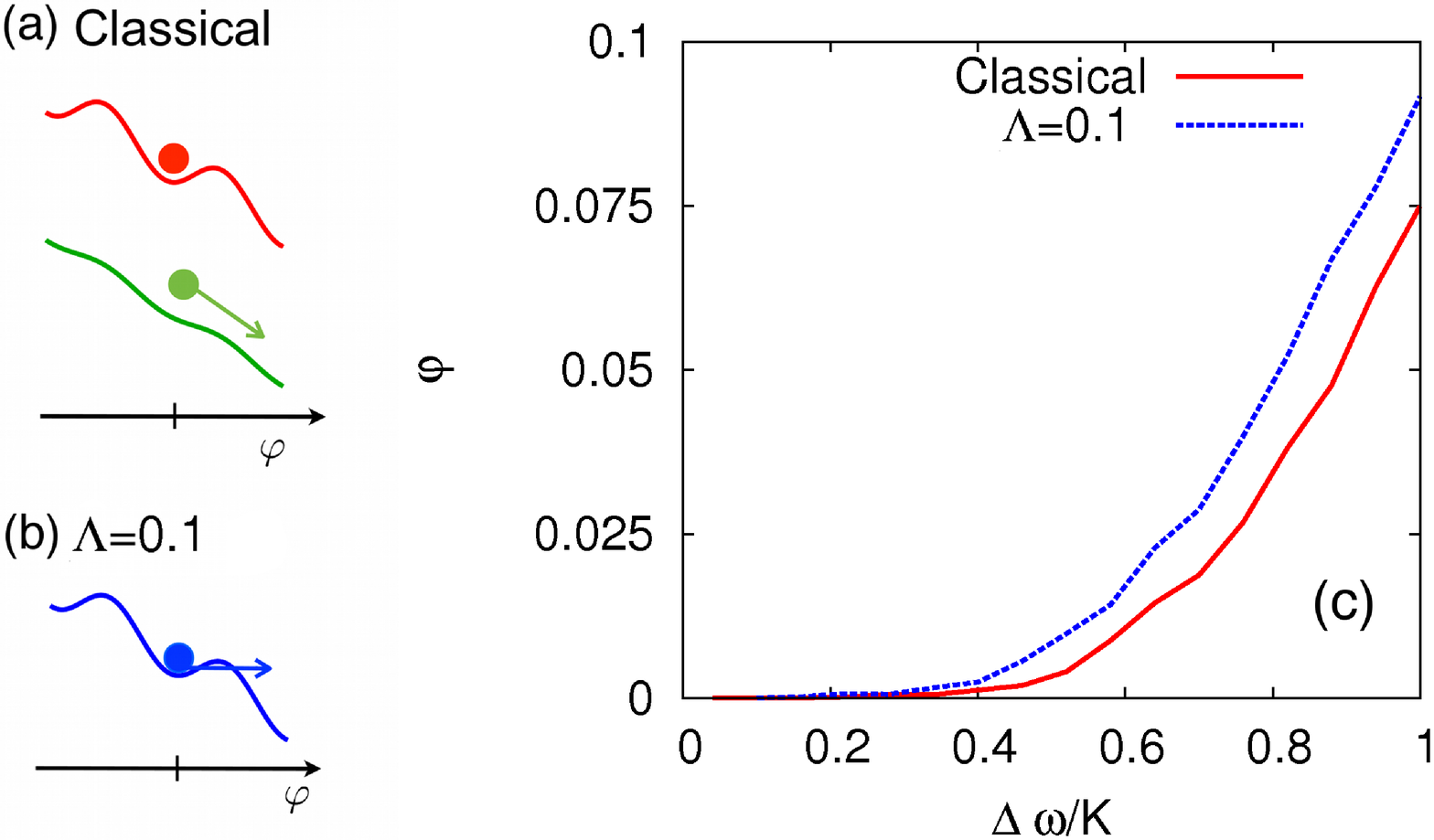}
\caption{(color online) {\bf System of two coupled Kuramoto oscillators}. 
The top-left part of the figure shows the 
analogy between the system of two coupled oscillators and an overdamped particle in a washboard 
potential. 
The two possible regimes are shown: synchronized state (the particle is at rest $\dot \varphi=0$ at a 
local minimum) and unsynchronized phase (the particle drifts across the potential). 
Below we illustrate the possibility that tunnelling provides to anticipate the 
drifting state. 
On the right part we show the result of the computation of the velocity $\dot \varphi$ as a function of $\Delta \omega/K$ 
for the classical (solid red line) and quantum (dashed blue line) systems. }
\label{fig:mechanical}
\end{figure*}
Figure \ref{fig:2}.a shows
the typical synchronization diagram, namely, the value of $r$ as a function 
of the coupling strength $K$.  
The comparison of the semiclassical (for $\Lambda=0.1$) and classical curves $r(K)$ 
evinces that quantum fluctuations delay the onset of synchronization, {\it i.e.}, 
the critical point $K_{\mathrm c}$ is seen to move to larger values with $\Lambda$.
We have also considered the evolution for the distribution of the phases as a function of $K$ 
to monitor the microscopic fingerprint of the synchronization transition. 
The evolution of the classical and quantum Kuramoto models is shown in figures 
\ref{fig:2}.b and \ref{fig:2}.c, respectively.

To explain the delay in the synchronization onset introduced by quantum fluctuations 
we resort to the simplest situation: two coupled Kuramoto oscillators. 
In this case the Kuramoto model (\ref{langevin}) consists of just two coupled equations for the evolution of $\theta_1$ and $\theta_2$.  
By taking the difference of those two equations and introducing as a new variable the phase difference, $\varphi := \theta_1 - \theta_2$, we obtain for its evolution the following equation: 
\begin{equation}
\dot \varphi = \Delta \omega  - K \sin \varphi
+
2 D \xi
\label{particle}
\end{equation} 
The latter equation describes the evolution of an overdamped particle in a washboard potential  
(see Figure \ref{fig:mechanical}). 
With this image in mind, we map the synchronous
movement of the two oscillators (defined as a state in which the frequencies of the oscillators 
are locked: $\dot \theta_1 = \dot \theta_2$) with the resting state of the overdamped particle 
inside a local minimum of the potential energy ($\dot \varphi = 0$).  
On the other hand, when the two oscillators are not synchronized the particle drifts across the 
potential ($\dot \varphi \neq 0$). 
Both situations are shown in Fig. \ref{fig:mechanical}.

The quantum version for the diffusion of an overdamped particle in a 
periodic potential has been previously studied in Ref.~\cite{uczka2005}. 
The main result is that the scape rate of the particle, and thus its unlocking mechanism, 
is enhanced through quantum fluctuations. 
This effect can be seen as a consequence of the enhancement of the transition probability 
for energies below the height of the barrier which is nothing but the well-known tunnel effect 
\cite{Garcia-Palacios2004}. 
In Fig. \ref{fig:mechanical} we show, for both the classical and semiclassical ($\Lambda=0.1$) 
systems of two coupled Kuramoto oscillators, the value of $\dot \varphi = 0$ as a function 
of the ratio between the difference of the natural frequencies of the two oscillators $|\Delta\omega|$ 
and the coupling $K$. 
It is clear that, as stated above, quantum tunnelling facilitates the drift or, equivalently, delays
 the transition to the synchronous state.

\section{Analytical expression for the synchronization onset}
\label{sec:K}

Coming back to the original model of $N$ interacting oscillators,
we now make an analytical estimation of the value for critical coupling at which the 
synchronization transition occurs. 
The procedure is a generalization of the one presented in Ref.~\cite{Sakaguchi1988} and 
takes advantage of the mean field description of the Kuramoto 
model. 
The derivation (detailed in  Appendix \ref{app:Kc}) yields a rather simple equation 
for the critical coupling: 
\begin{equation}
\label{kqkcl}
K^{\rm q}_{\mathrm c}= (1+ \Lambda)K_{\mathrm c}\;,
\end{equation}
being $K_{\mathrm c}$ the classical critical value shown in Eq. (\ref{KCS}).

The above result states  that quantum fluctuations act by effectively 
decreasing the coupling strength with the degree of quantumness $\Lambda$. 
Coming back to the physical image of a particle in a washboard potential, we 
can consider the effect of the quantum correction by considering the first and 
third terms in the right hand side of equation (\ref{q-l}). 
In this way, quantum corrections can be casted in the form of an effective potential:
\begin{equation}
V_{\rm eff} = V + \Lambda V^{\prime \prime},
\end{equation}
that in the particular case of the washboard potential reads: 
\begin{equation}
V_{\rm eff} = -\Delta \omega \;\varphi  - (K - \Lambda) \cos \varphi\;.
\end{equation}  
The above equation makes clear that tunnelling is formally reflected by an effective 
barrier reduction that yields the observed shift to higher values for the critical coupling. 

\begin{figure*}
\centering
\includegraphics[width=0.98\textwidth, angle=0]{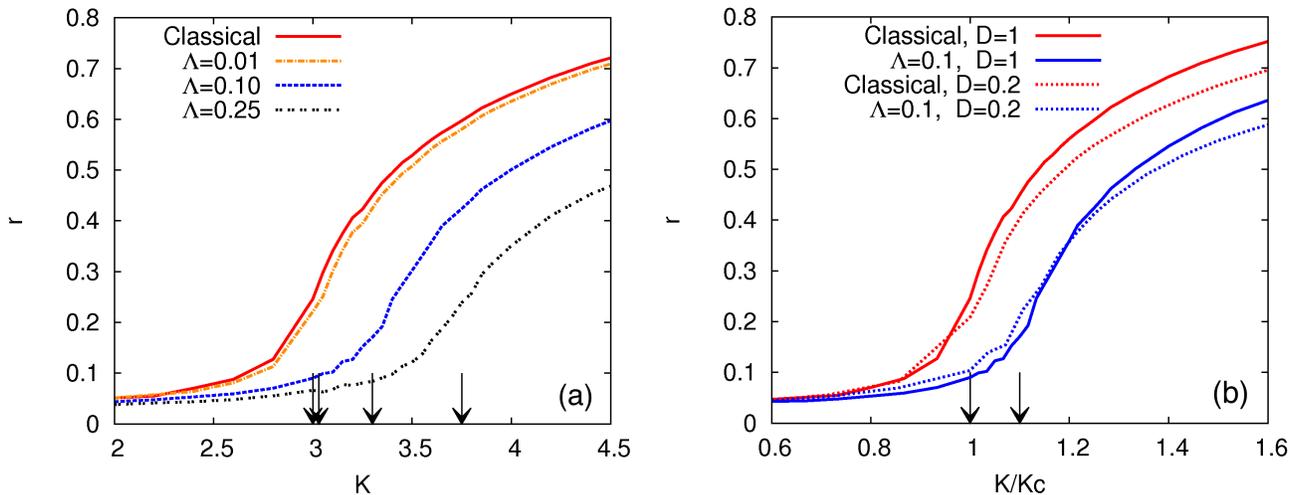}
\caption{(color online) {\bf Analysis of the behavior of the critical coupling}. 
In panel (a) we show the synchronization diagrams $r(K)$ for different values of the degree 
of quantumness $\Lambda$. 
Note that the coupling strength has been normalized to its value in the classical ($\Lambda=0$) limit. 
The analytical estimation in equation (\ref{kqkcl}) is shown by the vertical lines 
confirming its validity. 
In panel (b) we plot $r(K)$ for two different temperatures, corresponding to $D=1$ and $D=0.2$, 
for the classical and quantum ($\Lambda=0.1$) models. 
Again the analytical estimation is shown by the vertical lines.}
\label{fig:4}
\end{figure*}

Our analytical estimation for $K_{\mathrm c}^{\mathrm q}(\Lambda)$ is
plotted in figure \ref{fig:2}  and \ref{fig:4}
(vertical arrows) confirming its validity. 
To corroborate further the correctness of equation (\ref{kqkcl}), we explore the 
synchronization transition for different values of $\Lambda$ in figure \ref{fig:4}.a.  
As expected the onset of synchronization shifts to higher values as the degree of 
quantumness increases. 
Again, the predicted value for $K_{\mathrm{c}}^{\mathrm{q}}$ is plotted (vertical arrows) 
corroborating the validity of equation (\ref{kqkcl}). 

 To complete our study, we show in  \ref{fig:4}.b  the dependence of the 
synchronization diagram with the {\it thermal fluctuations}, $D$, both for the classical 
and quantum ($\Lambda=0.1$) cases. 
In all the curves explored the coupling $K$ is rescaled by the corresponding critical 
coupling $K_{\mathrm c}$ in the classical regime. In this way we show both for the 
classical and quantum cases, the robustness of the critical value (\ref{kqkcl}) against 
temperature changes.

\section{Discussion}
\label{sec:conc}

The search for quantum corrections to classical phenomena has been pervasive in physics.
Some examples related to our work are the generalization to the
quantum domain of chaos \cite{Ott}, dissipation \cite{Weiss}, 
random walks \cite{SanchezBurillo2012}, etc. 
Each of these examples finds its own difficulties when incorporating quantum 
fluctuations and unveiling their role. 
Some of these obstacles are the quantum linearity versus the typical non-linearity 
of classical systems and the quantization of non-Hamiltonian system or phenomenological equations. 
Overcoming these obstacles provides with a consistent quantum description that opens the quantum door 
to a variety of classical problems and their associated physical phenomena.

Among the most studied phenomena in (classical) complex systems is synchronization. This emergent phenomena is as intriguing as beautiful, since it covers from the description of the {\it sympathy} of clocks to the neuronal functioning in our brain, thus overcoming the disparately diversity in the spatial and time scales associated to the bunch of systems in which synchronization is observed. However, the concept of synchronization was usually associated to the classical domain as the typical examples of  clocks, fireflies or humans are too macroscopic to think about the need of introducing quantum fluctuations in the description of the associated dynamical models.

Recently, some experimental works have shown that synchronization can be observed in the lab within Josepshon Junction arrays \cite{Wiesenfeld1998}, nanomechanical \cite{Matheny2013} or optomechanical systems \cite{Heinrich2011}. All of these systems share one prominent property: they behave quantum mechanically at sufficiently low temperatures. Therefore, adapting the concept of synchronization among coupled entities within the quantum theory is, apart from an interesting theoretical issue, a must imposed by  the rapid experimental advances.

A first step consists in taking the most widely used framework for studying synchronization phenomena, 
the Kuramoto model, and adapting it to the quatum domain. Being a paradigmatic theoretical setup, the quantization of the Kuramoto model opens the door to the theoretical study of quantum synchronization in the widest possible manner. 
To this end, and to overcome the non-Hamiltonian character of the Kuramoto equations, 
we have mapped the model to an overdamped Langevin equation which has a Hamiltonian description by 
embedding the system in a bath of oscillators. 
In this way, the quantization of the Kuramoto model is straightforward and it includes its classical counterpart 
as a limiting case: the quantum version incorporates quantum fluctuations  for the phases while the strength 
of these quantum corrections are encoded in a single parameter.

The route chosen here must be understood as complementary to the study of particular models of coupled quantum systems. The reason is twofold. First, we aim to be as general as possible. The essence of an emergent phenomena is its ability of describe very different situations with different microscopic dynamics. This is the goal of the Kuramoto model, as it explains the synchronization without resorting to the specific dynamics. Second, a force brute study of many body quantum entities is a very difficult task that usually implies the reduction of the system to a few coupled systems. However, the observation of a true synchronization transition demands hundreds or thousands of interacting dynamical systems.   

Being general, the results obtained allow to make general statements about 
the impact that quantumness has on the synchronization of coupled dynamical units. 
The most important one is that quantum fluctuations delay the appearance of a synchronized state. 
The explanation of this effect relies on the fact that in the quantum domain the phases not only have 
a different natural frequency but also the fluctuations around the classical trajectories are different 
depending on those internal rhythms. 
To illustrate this interpretation we recall the simple case of
two coupled Kuramoto oscillators. In this case quantum fluctuations are nothing but
thermal assisted tunneling favoring the phase unlocking.
Therefore, the coupling needed to synchronize the two oscillators is higher 
in the quantum limit.   

Finally, we want to point out that in a recent publication the question about synchronization in quantum 
evolutions was also discussed \cite{Mari2013}. Under rather general conditions they find bounds for the degree of synchronization based on the Heisenberg uncertainty principle: the phases, derived as averages of non-conmuting operators, cannot take values infinitely close. 
Instead, in our case, focused on the quantum version of the Kuramoto model, we have discussed, not the maximum degree of synchronization but the critical onset for the  appearance of partially synchronized states. 
In this case quantumness also limits the emergence of a synchronous state. 
Therefore, pretty much like in what happens in quantum chaos, synchronization seems to be a  quasi-classical 
phenomena \cite{Liu2013}.


\begin{acknowledgments}
This work has been partially supported by the Spanish MINECO under projects FIS2011-14539-E (EXPLORA program) FIS2011-25167 and FIS2012-38266-C02-01, by the Comunidad de Arag\'on (Grupo FENOL). J.G.G. is supported by MINECO through the Ram\'on y Cajal program.
LAP was supported by CODI of Universidad de Antioquia under contract number E01651 
and under the \textit{Estrategia de Sostenibilidad 2013-2014} and by COLCIENCIAS of Colombia 
under the grant number 111556934912.
\end{acknowledgments}


\appendix

\section{The semiclassical Kuramoto model: technical details}
\label{app:q-K}

In this  appendix we  provide some technical details for obtaining the
semiclassical Kuramato model in Eq.~\eqref{q-l}. As sketeched in the main
text, See \ref{sect:sketch},  we need: the equilibrium density matrix,
 calculate some coefficients in a Fokker-Planck equation and
transform the latter to a Langevin type equation.

\subsection{Equilibrium Density Matrix: Path Integral Formalism}
\label{app:eq-pi}

Let us compute the equilibrium
density matrix.  In particular we are interested in the reduced
density matrix (at equilibrium):
\begin{equation}
\varrho_\beta = {\rm Tr_{bath}} \{ 
W_\beta
\},
\end{equation}
where $W_\beta$ is the total equilibrium density operator, $W_\beta
\sim {\rm e} ^{-\beta (H_{\rm sys}
+
H_{\rm bath}
+
H_{\rm int} ) }$.
The equilibrium reduced density 
matrix can be expressed as \cite{Ingold2002}
\begin{align}
\label{equ:EffAct}
\varrho_\beta
({\boldsymbol \theta}, {\boldsymbol \theta}^{\prime})
=
\frac{1}{\mathcal{Z}}
\int_{\theta_1}^{\theta_1^\prime}
{\mathcal D} \theta_1
\cdot \cdot \cdot
\int_{\theta_N}^{\theta_N^\prime}
{\mathcal D} \theta_N
\;
{\rm e}^{- \frac{1}{\hbar} S^E_{\rm eff} [{\boldsymbol \theta}]},
\end{align}
with the effective action
\begin{align}
\nonumber
S^E_{\rm eff} [{\boldsymbol x}]
= &
\int^{\hbar \beta}_0 {\rm d} \tau
\, \Big ( 
\sum_j \frac{1}{2} m \dot \theta_j^2
+
V(\theta_1, ..., \theta_N)
\, \Big )
\\ 
&+
\frac{1}{2} \sum_j \int^{\hbar \beta}_0 {\rm d} \tau 
\int^{\hbar \beta}_0 {\rm d} \sigma 
K(\tau - \sigma) \theta_j (\tau) \theta_j(\sigma),
\end{align}
which contains the kernel
\begin{equation}
K (\tau) =
\frac{m}{\hbar \beta}
\sum_n
| \nu_n |
\hat \gamma
( | \nu_n | )
{\rm e}^{i \nu_n \tau},
\end{equation}
being $\nu_n$ the Matsubara frequencies,
\begin{equation}
\nu_n = \frac{ 2 \pi n}{\hbar \beta}.
\end{equation}
and the Laplace transform of the damping kernel is given by:
\begin{equation}
\hat \gamma (z) = \frac{2}{m}
\int_0^\infty
\frac{{\rm d} \omega}{\pi}
\frac{J(\omega)}{\omega}
\frac{z}{z^2 + \omega^2}.
\end{equation}

\subsection{Overdamped Equilibrium}

Based on previous works \cite{Ankerhold2005, Maier2010}
for the single particle case, we compute the equilibrium distribution
in the overdamped limit.
the overdamped dynamics refer to a regime
in the parameter space where damping is sufficiently strong to suppress 
the non-diagonal elements, \textit{coherences}, of the reduced density matrix, 
i.e., a regime where 
$\langle \theta_1, ..., \theta_N | \varrho_\beta | \theta_1^\prime,
..., \theta_N^\prime \rangle \sim \prod \delta (\theta_i - \theta_i^\prime)$.
These {\it semiclassical} diagonal contributions can be computed perturbatively
on the quantum fluctuations.

\subsubsection{Minimal path}
Let us denote the minimal action (ma) path as 
$x^{\rm ma}_i  \equiv \bar \theta_i$.
Besides, since we are interested in the diagonal contributions in the imaginary-time 
path integral in equation~(\ref{equ:EffAct}), this means for us to take the trajectories with 
\begin{equation}
\label{per-diag}
\bar \theta_i (0) = \bar \theta_i (\hbar \beta)
\equiv \theta_i,
\end{equation}
{\it i.e.}, periodic trajectories with frequencies $\nu_n$.
The minimal action path satisfies the generalized Lagrange equations \cite{Grabert1988}
\begin{equation}
\label{min-path}
m \ddot {\bar \theta}_i
  - \frac{\partial V}{ \partial \bar \theta_i}
-
\int_0^{\hbar \beta} {\rm d} \sigma
k(\tau - \sigma)  \bar \theta_i (\sigma) 
= 0.
\end{equation}
The periodic condition in equation~(\ref{per-diag}) suggests to Fourier expand ${\bar \theta}_i(\tau)$,
such that
\begin{equation}
\bar \theta_i(\tau) = \sum_n \theta_{n, i} {\rm e}^{i \nu_n \tau},
\end{equation} 
where the Fourier components satisfy
\begin{equation}
- \nu_n ^2 \theta_{n,i} + \gamma (\nu_n ) \theta_{n,i} 
+
v_{n,i} = b_i,
\end{equation}
with
\begin{equation}
v_{n,i} = 
\int_0 ^{\hbar \beta}
{\rm d} \tau
\frac {\partial V}{\partial \theta_i} {\rm e}^{-i \nu_n \tau}
\end{equation}
and the inhomogenous term
\begin{equation}
b_i= \dot {\bar \theta}_i (\hbar \beta) - \dot {\bar \theta}_i (0),
\end{equation}
comes from the jumps and cups singularities arising from fact that the 
Fourier series expansion for $ {\bar \theta}_i(\tau)$ periodically continues 
the path outside the interval $0\le \tau\le \hbar \beta$ \cite{Grabert1988}.
Note that terms like $a_i= {\bar \theta}_i (\hbar \beta) - {\bar \theta}_i (0)$
are, in general, expected.
However, since we are interested in the 
diagonal contributions, they do not contribute to the present case.

At this point, we first notice that by making $n=0$ for $b_i$ we obtain
\begin{equation}
\label{bi}
b_i = \frac{\hbar \beta}{m} \frac {\partial V}{\partial \theta_i}.
\end{equation} 
Besides, the components $\theta_{n,i}$ with $n \neq 0$,
\begin{equation}
\label{thetani}
\theta_{n,i} = \frac{-b_i}{\nu_n^2 + \gamma (\nu_n )},
\end{equation}
are suppressed by dissipation. 
Hence
\begin{equation}
\label{theta0i-rel}
\theta_{0,i} \cong \bar {\theta}_i (0) + \frac{b_i}{\hbar} \Lambda,
\end{equation}
where $\Lambda$ measures the quantumness:
\begin{align}
\Lambda = &\frac{2}{ m \beta} \sum_n \frac{1}{\nu_n^2 + \gamma \nu_n}
\\ \nonumber
=&
\frac{\hbar}{m \pi \gamma}
\left ( \Psi \left [
\frac{\hbar \beta \gamma}{2 \pi}
\right]
-\mathrm{C} + \frac{2 \pi}{\hbar \beta \gamma} \right ),
\end{align}
being $\mathrm{C}= 0.577...$ the Euler-Mascheroni constant.
Note that in the limit $\hbar \to 0$, $\Lambda \to 0$, as it must be.
Thus, recovering the classical result.

The contribution of the minimal action can be further simplified by 
considering that
\begin{equation}
\label{equ:auxparint}
\frac{1}{2} \int {\rm d} \tau\,  \dot {\bar \theta}_i ^2  =  \frac{1}{2} \left
  [ 
\theta_i (\dot {\bar \theta}_i (\hbar \beta) 
-
\dot {\bar \theta}_i (0) )- \int {\rm d} \tau\, \theta_i \ddot {\bar \theta}_i 
\right ]
\end{equation}
together with (\ref{per-diag})
and replacing equation~(\ref{min-path}) in the second term at the r.h.s
of equation~(\ref{equ:auxparint}), such that
\begin{equation}
S = 
\frac{1}{2}
\sum_i
\theta_i b_i + \int_0^{\hbar \beta} {\rm d} \tau
\;
\big ( V 
- \frac{1}{2} \sum_i \bar {\theta}_i \partial_{\theta_i} V
\big ) \, .
\end{equation}
By using the relation (\ref{bi}) and by noticing that $\bar {\theta}_{i} \cong \theta_{0,i}$ 
[$\theta_{n,i}$ are suppressed, see equation~(\ref{thetani})], we have that 
$\bar {\theta}_{i} - \theta_i = b_i \Lambda /\hbar$ [Cf. equation~(\ref{theta0i-rel})].
Hence,
\begin{equation}
S_{\rm ma}
=
\hbar \beta
V
-
\frac{1}{2}
\sum_i \hbar \beta^2 \Lambda (\partial_{\theta_i} V)^2.
\end{equation}

\subsubsection{Fluctuations around the minimal action path}
We study now the fluctuation around the minimal path
\begin{equation}
\theta_i = \bar {\theta}_i + y_i,
\end{equation}
subjected to the boundary conditions:
\begin{equation}
\label{bc-y}
y_i(0)= y_i (\hbar \beta) = 0.
\end{equation}
Consequently, the correction to the path integral reads,
\begin{equation}
F(q) =
\int {\mathcal D} y_1 
\cdot \cdot \cdot 
\int {\mathcal D} y_N
{\rm e}^{-1/\hbar \int _0 ^{\hbar \beta} {\rm d} \tau
\langle y |  \mathsf{L} | y \rangle},
\end{equation}
where we have used an economical notation, {\it  {\`a} la Dirac}, for
the quadratic form $\langle y | \mathsf{L} | y \rangle = \sum L_{ij} y_i
y_j$, being $\mathsf{L} = \{\{L_{ij}\}\}$ defined as
\begin{equation}
 \mathsf{L} = - \mathsf{I}
\left ( m \frac{d^2}{d \tau^2} 
+
\int_0^{\hbar \beta} {\rm d} \sigma\, k(\tau - \sigma) 
\right )
+
 \mathsf{V}^{\prime \prime},
\end{equation}
where $\mathsf{I}$ is the identity matrix and  the 
second-derivative-potential-matrix $\mathsf{V}^{\prime \prime} = \{\{V_{ij}^{\prime \prime} \}\}$ 
is defined as,
\begin{equation}
 V^{\prime \prime} _{ij} :=
\frac {\partial V}{\partial \theta_i \partial \theta_j}.
\end{equation}

We proceed as above and Fourier expand the fluctuations 
around the minimal path [Cf. equation~(\ref{bc-y})],
\begin{equation}
y_i = \frac{1}{\hbar \beta} \sum_n y_{n,i} {\rm e}^{i \nu_n \tau},
\end{equation}
which allows us to effectively replace the boundary condition $y_i(0)=0$ 
in terms of a product of Dirac's delta functions, $\prod_i \delta [y_i(0)]=\prod_i \delta
[1/\hbar \beta \sum _n y_{n,i}]$, in the integral expressions above, i.e., 
by changing
\begin{equation}
\prod_i \delta
[y_{n,i}]
\sim \int 
\prod_i {\rm d} \mu_i 
{\rm e}^{i/\hbar \beta \langle \mu | y_n \rangle}
\end{equation}
where $\langle \mu | y_n \rangle = \sum_i \mu_i y_{n,i}$.
Therefore,
\begin{equation}
F(q)  \sim \int \prod_i {\rm d} \mu_i 
\prod_n \prod_j
{\rm d} y_{n,j}
{\rm e}^{i/\hbar \beta \langle \mu | y_n \rangle}
{\rm e}^{-1/\hbar \beta \langle y_n |  \mathsf{A}_n | y_n\rangle},
\end{equation}
with,
\begin{equation}
\label{An}
 \mathsf{A}_n =  \mathsf{I} \lambda_n  +  \mathsf{V}^{\prime \prime}
\;
\quad \mathrm{and}\quad
\lambda_n = \nu_n^2 + |\nu_n | \gamma.
\end{equation}

This is a Gaussian integral that can be performed by resorting twice to the
formula  
\begin{equation}
\int
\prod_j {\rm d} \theta_j
{\rm e} ^{- \langle x |  \mathsf{A}| x \rangle +  \langle b | x \rangle }
= \sqrt { \frac {\pi^N} {{\rm det}  \,  \mathsf{A}} }
\; 
{\rm e} ^{- \langle b |  \mathsf{A}^{-1} | b \rangle}.
\end{equation}
So that
\begin{equation}
F(q)
\sim
\sqrt { \frac{ \prod_n {\rm det}  \,  \mathsf{A}_n^{-1}}{\sum _n {\rm det}  \,  \mathsf{A}_n^{-1} }}.
\end{equation}
%
Up to first order in $1/\gamma$, we get [Cf. equation~(\ref{An}]:
\begin{equation}
 \mathsf{A}_n^{-1}
\cong 
\frac{1}{\lambda_n}
 \mathsf{I}-
\frac{1}{\lambda_n^2}
 \mathsf{V}^{\prime \prime}.
\end{equation}
To be consistent, we also need to compute the determinants at first 
order in $1/\gamma$ \cite{detapprox}
\begin{equation}
{\rm det} \,  \mathsf{A}_n ^{-1}
\cong 
\frac{1}{\lambda_n^N}
{\rm e}^{-{\rm Tr}[  \mathsf{V}^{\prime \prime}]/\lambda_n}.
\end{equation}

Based on all the consideration above, in the next appendix we explicitly 
present the thermal equilibrium state with first order corrections in the 
fluctuations along the semiclassical minimal path results and derive
the associated Smoluchowski equation.

Based on the result obtained in section \ref{app:eq-pi}, the equilibrium
density matrix in the overdamped limit reads, see also Eq.~(\ref{rho-eq}):
\begin{equation}
\label{rho-eq-app}
 P_\beta (\theta)
=
\frac{1}{\mathcal Z}
{\rm e} ^{ -\beta \Lambda  \sum_i 
V_{i,i}^{\prime \prime}
}
\,
{\rm e}^{-\beta V +
\frac{1}{2}
 \beta^2 \Lambda \sum_i (
V_i^\prime
)^2 }
\, .
\end{equation}
where we have introduced the notation $P_\beta (\theta)$. In the
overdamped limit only the diagonal elements  $\varrho_\beta (\theta,\theta)  $
matter [Cf. Eq. (\ref{P-def})].  We have also introduced the compact
notation [see the main text, below Eq. \eqref{q-l}]: $V_{i, ..., k}^{\prime
  ... \prime} \equiv \partial_{\theta_i, ..., \theta_k} V$.
%

\subsection {The Quantum Master Equation for the  Kuramoto Model: q-K}
We proceed here as Ankerhold et al. in Refs.~\cite{Ankerhold2005,
Maier2010}:

\subsubsection{One-Particle Master Equation}
As a warm up, let us consider the one-particle model.
In the classical case, the Fokker-Planck equation can be 
expressed as
\begin{equation}
\label{eq:s01}
\partial_tP=\partial_{\theta}LP 
\end{equation}
where
\begin{equation}\label{eq:s02}
L=D_1({\theta})+\partial_{\theta}D_2
\end{equation}
with,
\begin{equation}
D_1=V'=\partial_{\theta}V
\end{equation}
and
\begin{equation}
D_2=\frac{D}{\gamma^2}=\frac{k_{\beta}T}{m\gamma}=\frac{1}{m\gamma\beta}=\frac{1}{\Gamma\beta},
\end{equation}
here $\Gamma := m\gamma$. 

Let us switch into the quantum regime.
The reduced density matrix for the single 
particle case, [See~(\ref{rho-eq})]  reads:
\begin{equation}
\label{eq:s03}
P_{\beta}=\frac{1}{Z}e^{-{\beta}{\Lambda}V''}
e^{(-{\beta}V+\frac{\beta^2\Lambda}{2}V'^2)},
\end{equation}
where $Z$ is the partition function and
\begin{equation}
S=-{\beta}V+\frac{\beta^2\Lambda}{2}V'^2.
\end{equation}
Up to leading order in $\Lambda$,
\begin{equation}
\label{equ:OpPbeta}
P_{\beta}=\frac{1}{Z}\left(1-{\beta}{\Lambda}V''\right)
e^{-{\beta}V}\left(1+\frac{\beta^2\Lambda}{2}V'^2\right).
\end{equation}

Imposing the consistency condition $\mathcal L  \varrho_\beta = 0$ together with the
election for 
$D_2$:
\begin{equation}
D_2=\frac{D}{\gamma^2}(1+{\beta}{\Lambda}V'')=\frac{D}{\gamma^2F},
\end{equation}
where $F=1-{\beta}{\Lambda}V''$, we find

\begin{equation}
D_1=\frac{D}{\gamma^2}\beta{V'}=\frac{1}{m\gamma}V'.
\end{equation}
This yields the QME for the single case in the overdamped limit:
\begin{equation}
\label{single}
\partial_tP=\partial_{x} \left\{  \frac{1}{m\gamma}V'
+\partial_{x}\left[ \left(\frac{D}{\gamma^2}(1+{\beta}{\Lambda}V''\right)\right] \right\} P.
\end{equation}
$\bigskip$

\subsubsection{$N$-Particles Master Equation}

The generalization for (\ref{eq:s01}) and(\ref{eq:s02}) for the
multivariate case reads:
\begin{equation}
L=D_{1,i}({\theta})+\partial_{{\theta},i}D_{2,i},
\end{equation}
whereas the stationary solution $P_{\beta}$ in equation~(\ref{rho-eq}) can be
rewritten as,
\begin{equation}
\label{StSl:nP}
P_{\beta}=\frac{1}{Z}e^{-{\beta}{\Lambda}\mathrm{Tr}(\mathsf{V}^{\prime\prime})}
e^{-{\beta}V+\frac{\beta^2\Lambda}{2}\mathbf{V}^{\prime}\cdot\mathbf{V}^{\prime}},
\end{equation}
where $\mathrm{Tr}(\mathsf{V}^{\prime\prime})$ denotes trace of the matrix 
$\mathsf{V}^{\prime\prime}$. 
The stationary solution can be always be written as 
\begin{equation}
\label{Pformal}
P_\beta (\theta) \equiv
\frac{1}{Z}F({\theta})e^S.
\end{equation}
%
%
With the experience gained in the single particle case,  our election for $F$
and ${\rm e}^S$ will determine the values for $D_{1,i}$ and $D_{2,i}$.
If we choose $F=1$ we do not recover the overdamped equation for  a
quantum harmonic oscillator in the one-particle limit, wich is an
exact result \cite{Maier2010}. 
On the other hand we can set, by analogy with the single site case,  $F=
e^{-{\beta}{\Lambda}\mathrm{Tr}(\mathsf{V}'')}$.  For recovering the
uncoupled case,  $F$ can be rewritten as
 $F = \prod_i F_i$. The actual value for  $F_i$ must recover the
 master equation for the
 harmonic oscillator.

We choose
\begin{equation}
\label{D2-app}
D_{2,i}=\frac{D}{\gamma^2F_i}
\end{equation}
and impose $P_{\beta}$, Eq. \eqref{Pformal},  to be the stationary solution:
\begin{widetext}
\begin{align}
\nonumber
D_{1,i}P_{\beta}+\partial_{{\theta}_i}D_{2,i}P_{\beta} &=
\\ \nonumber D_{1,i}\prod_jF_je^S
+\frac{D}{\gamma^2}\partial_{{\theta}_i}\left(\prod_{j\neq{i}}F_j\right) e^S&=
\\ \nonumber
{\rm e}^S
\left [D_{1,i}\prod_jF_j+\frac{D}{\gamma^2}\left(\prod_{j\neq{i}}F_j\right)\partial_{{\theta}_i}S 
+\frac{D}{\gamma^2}\sum_{j\neq{i}}\left(F'_{j,i}\prod_{k\neq{j}\neq{i}}F_k\right)
\right ]
&=0 \;.
\end{align}
Thus,
\begin{equation}
\label{D1-app}
D_{1,i}=-\frac{D}{\gamma^2F_i}\partial_{{\theta}_i}S
-\sum_{j\neq{i}}\frac{D}{\gamma^2F_i}\frac{F'_{j,i}}{F_j}.
\end{equation}
We have already justified the form for $F$, giving
\begin{equation}
\label{Fapp}
F=\prod_i{F_i}=\prod_i{e^{-{\beta}{\Lambda}V^{\prime\prime}_{ii}}}.
\end{equation}
Collecting \eqref{D2-app}, \eqref{D1-app} and \eqref{Fapp} the final
form for the master equation is obtained describing a system of $N$
particles in the Smoluchowski regime:
\begin{align}
\label{eq:s04}
\partial_tP 
= \sum_i\frac{{\partial}}{{\partial}{\theta}_i}
\Big\lbrace  
\Big[ 
\frac{V'_i}{{\Gamma}F_i}-\frac{\beta}{{\Gamma}F_i}{\Lambda}
{\sum_j}
\frac{{\partial}V}{{\partial}{\theta}_j}
\frac{\partial^2V}{{\partial}{\theta}_i{\partial}{\theta}_j}
 +\frac{\Lambda}{{\Gamma}F_i}
\sum_{j\neq{i}}V^{\prime\prime\prime}_{jji}\Big]  
+ \frac{{\partial}}{{\partial}{\theta}_i}\left[ 
\frac{D}{\gamma^2F_i}\right] \Big\rbrace P.
\end{align}
It is easy to check that making $N=1$ above the single particle master
equation  \eqref{single} is recovered.
\end{widetext}

\subsubsection{The Langevin Equation}
Once we have derived the master equation, we can easily find the associated 
Langevin equation, in the form 
\begin{equation}
\frac{{\partial}{\theta}_i}{{\partial}t} = A_i(\boldsymbol{\theta},t)
+\sum_kB_{ik}(\boldsymbol{\theta},t)\xi_k(t),
\end{equation}
following the guidelines explained in Ref.~\cite{Garcia-Palacios2007} and Chap.~3 in Ref.~\cite{Risken1989}.
Here $\xi_k$ is Gaussian $\delta$-correlated white noise with
zero mean and variance $2D$.  Following
Ref.~\cite{Garcia-Palacios2007} the Langevin equation is equivalent to
the Fokker-Planck type equation for the probability distribution:
\begin{align}
\label{FPAB}
\partial_tP = &-\sum_i\frac{{\partial}}{{\partial}{\theta}_i}\left\lbrace
  \left[
    A_i+D\sum_{jk}B_{jk}\frac{{\partial}B_{ik}}{{\partial}{\theta}_j}\right]
  P \right\rbrace 
\\ \nonumber
&+
 D\sum_{ij}\frac{\partial^2}{{\partial}{\theta}_i{\partial}{\theta}_j}
 \left\lbrace \left[ \sum_k B_{ik}B_{jk} \right] P\right\rbrace.
\end{align}
Comparing \eqref{eq:s04} and \eqref{FPAB} the coefficients $A_i$ and
$B_{ij}$ can be identified.  For the concrete case of the {\it
  Kuramoto Potential} \eqref{eff-pot} we finally endup in the semiclassical
Kuramoto model in Eq.~\eqref{q-l}.
%


\section{Critical coupling value}
\label{app:Kc}
We generalize here the work presented in Ref.~\cite{Sakaguchi1988} to the
quantum domain.

\subsection{Periodicity and self-consistency of the master equation}

The order parameter $r$ is given by:
\begin{equation}
re^{i(\omega_0t+\phi_0)}=\frac{1}{N}\sum^N_{j=1}{e^{i\phi_j}}.
\end{equation}
The Kuramoto potential $V$ (\ref{eff-pot}) in a mean-field
approximation reads:
\begin{equation}
\label{Vofpsi}
V = -\omega\psi-Kr\cos{\psi} \, .
\end{equation}
Nor $V$, neither the stationary solution  (\ref{rho-eq}) are $2 \pi$-periodic.
We have to find a \textit{periodic} stationary solution.
Following a similar procedure as the one performed by Risken (see pgs. 98 and 287-288
in Ref.~\cite{Risken1989}), we derive the following periodic stationary solution 
\begin{align}
\label{StSl:P}
P(\psi;\omega)=&{\mathrm e}^{-\beta{V_{\mathrm{eff}}}}P(0;\omega)
\left[1+\frac{({\mathrm e}^{-2\beta\pi\omega}-1)
\int^\psi_0 \mathrm{d}\phi\, {\mathrm e}^{\beta{V_{\mathrm{eff}}}}}{\int^{2\pi}_0 \mathrm{d}\phi\,
{\mathrm e}^{\beta{V_{\mathrm{eff}}}}} \right],
\end{align}
with
$V_{\mathrm{eff}}=V-\frac{1}{2} \beta\Lambda V'^2 + \Lambda V''$.
In the classical limit $\Lambda \rightarrow 0$, 
$V_{\mathrm{eff}} \rightarrow V$, recovering the classical periodic 
stationary solution derived by Sakaguchi \cite{Sakaguchi1988}.
It is not hard to check that the $2 \pi$-periodic distribution
(\ref{StSl:P}) is also a   stationary solution for (\ref{eq:s04}).

\subsection{Critical value}

We follow  Sakaguchi \cite{Sakaguchi1988}  for finding
the critical coupling strength $K_{\mathrm c}^{\mathrm q}$. 
The order parameter $r$ can be expressed in terms of $\psi$ as:
\begin{equation}
\label{autocon1}
r=\int_{-\infty}^{\infty}\mathrm{d}\omega\, g(\omega) \int_0^{2\pi} \mathrm{d}\psi\, n(\psi;\omega)\mathrm{exp}(\mathrm{i}\psi).
\end{equation}
Replacing  (\ref{StSl:P}) above, we have a self-consistent equation for $r$. 
In the right hand of  (\ref{autocon1}), the imaginary part is always zero, because 
$g(\omega)$ is symmetric around $\omega=0$. 
The real part is expanded in powers of $Kr/D$,
\begin{widetext}
\begin{equation}
r=Kr\left[ \int_{-\infty}^{\infty}\mathrm{d}\omega\, g(\omega) 
\frac{\pi\omega / D [1+\Lambda(\omega^2 / D^2-1)]
[1+\coth(\pi\omega/D)]}{(\omega^2/D^2+1)} \right] 
+\mathcal{O}\left[\left( Kr/D\right)^2\right].
\end{equation}
Assuming a peaked $g(\omega)$-distribution,  we also expand
around $\omega = 0$, obtaining:
\begin{equation}
\label{autocon2}
r=Kr\left[ \int_{-\infty}^{\infty}\mathrm{d}\omega\,g(\omega) \frac{(1-\Lambda)(1+\pi\omega/D)}{(\omega^2/D^2+1)} \right] +
\mathcal{O}\left[\left( Kr / D \right)^2\right].
\end{equation}
Being $g(\omega)$  an even function, the linear term
$\pi \omega /D$ does not contribute to the integral.
Finally, the critical coupling strength, as a function of the temperature, is obtained from (\ref{autocon2}), 
\begin{equation}
K_{\mathrm c}^{\mathrm q}(\beta)=\frac{2}{(1-\Lambda)\int_{-\infty}^{\infty}\mathrm{d}\omega\,g(\omega)\frac{D^2}{(\omega^2+D^2)}}.
\end{equation}
As $K$ increases, a non-trivial solution branches off the trivial
solution $r=0$ at $K=K_{\mathrm c}$. 
This solution reduces to the classical one \cite{Strogatz2000,Sakaguchi1988} 
when $\Lambda=0$ at the classical critical coupling strength $K^{\mathrm c}_{\mathrm c}$.
A simple relation between the classical and the quantum critical
values can be obtained
\begin{equation}
K_{\mathrm c}^{\mathrm q}(K^{\mathrm c}_{\mathrm c};\Lambda)=\frac{K^{\mathrm c}_{\mathrm c} }{(1-\Lambda)}.
\end{equation}
\end{widetext}


\begin{thebibliography}{99}
\bibitem{detapprox} Here we have made use of the results derived in Ref.~\cite{Ipsen2011}.
In particular, let us define $ \mathsf{M} =  \mathsf{D} +  \mathsf{O}$, which can be
re-expressed as $  \mathsf{M} = {\mathsf D}({\mathsf I} + \mathsf{A} ) $ with 
$\mathsf{A} =  \mathsf{D}^{-1}  \mathsf{O}$.
The determinant of $\mathsf{M}$ reads 
$ {\rm det} \,  \mathsf{M} = \det \,  \mathsf{D} \; {\rm e}^{{\rm Tr} [ \log (  \mathsf{I}+  \mathsf{A}) ]}$,
with 
$\log (  \mathsf{I}+  \mathsf{A} ) =  \mathsf{A} - \frac{1}{2}  \mathsf{A}^2 + \frac{1}{3}  \mathsf{A}^3 -
\cdot \cdot  \cdot \; {\rm if}\, \rho( \mathsf{A}) < 1$.
\end{thebibliography}
\end{document}